\documentclass[11pt,a4paper]{article}
\usepackage{jheppub_kim}
 \usepackage{slashed}
\usepackage{longtable}
\usepackage{epsfig}
\usepackage{amsmath}
\usepackage{graphicx}
 \usepackage{graphics}
\usepackage[hang,nooneline,scriptsize]{subfigure}
\usepackage{array}
\begin{document}
\title{\Large Tsallis and Kaniadakis holographic dark energy with Complex Quintessence theory in Brans–Dicke cosmology}
\author[a]{J. Sadeghi,}
\author[b]{S. Noori Gashti,}
\author[a]{T. Azizi,}
\affiliation[a] {Department of Physics, Faculty of Basic Sciences, University of Mazandaran P. O. Box
47416-95447, Babolsar, Iran.}
\affiliation[a] {Canadian Quantum Research Center 204-3002 32 Avenue Vernon, British Columbia V1T 2L7 Canada.}
\affiliation[b] {Department of Physics, Faculty of Basic Sciences, University of Mazandaran P. O. Box 47416-95447, Babolsar, Iran.}
\emailAdd{pouriya@ipm.ir}
\emailAdd{saeed.noorigashti@stu.umz.ac.ir}
\emailAdd{t.azizi@umz.ac.ir}
\abstract{In this paper, by considering the complex form of the quintessence model, we study two different dynamic structures of holographic dark energy as Tsallis and Kaniadakis in the framework Brans-Dicke cosmology. In each setup, we employ non-interacting and interacting cases and calculate some cosmological parameters such as the equation of state $\omega$. We also discuss the $ \omega- \omega '$ behavior. By modifying the potential and studying the scalar field dynamics, we examine the complex quintessence cosmology.
	In addition, considering the two parts of the quintessence field effects, i.e., real and complex, and considering the fractional energy density $ \Omega_{D} $, we examine whether it can describe a real universe or not.
	We also specify that the fractional energy density can not be arbitrary between 0 and 1. In other words, it depends on the  Tsallis, Kaniadakis, and Brans-Dicke cosmology free parameters.
	We create a relationship between the fractional energy density and other parameters introduced in the text such as $\delta$, $b^{2}$, $\alpha$ and $\beta$ for each model separately. Finally, we compare the obtained results of models to each other and the latest observable data.\\\\
}

\maketitle
\section{Introduction}
As we know, one of the most common issues in modern cosmology is explaining the accelerating expansion of the universe. To describe the cosmic speedup, the researchers have benefited from various observations such as supernovae Ia, CMB anisotropies and large-scale structures \cite{1,2,3,4,5,6}.
Among the various theories that have been proposed for explaining this accelerated expansion, the dark energy scenario with a negative pressure has attracted a lot of attention.
Due to the unknown nature of dark energy, different models of dark energy have been introduced in the literature.
Some of these models are the cosmological constant, Chaplygin gas, interacting dark energy models,  ghost condensate, quintessence field, phantom field, quintom, tachyon models, braneworld models and  K-essence\cite{7, 8, 9,10,11,12,13,14,15,16,17,18}.

Another important approach that researchers have introduced to explain this unknown nature, is dark energy from a holographic point of view. Many studies have been done on this subject and the results are compared to the latest observational data to find the best model to explain the unknown nature of dark energy \cite{19,20,21}.
On the other hand, in generalized statistical mechanics, a new entropy is used in the concept of the black hole physics, which has led to the introduction of a new model in holographic dark energy, which is interpreted as Tsallis holographic dark energy \cite{22,23,24,25,26,27,28,29,30}.
Other models have been studied to explain the unknown nature of dark energy namely Kaniadakis, Barrow, and R$e$nyi  holographic dark energy\cite{31,32,33}.
As we know, one of the most important scalar models for describing dark energy is a scalar field called quintessence, which has been studied in two ways as real and complex scalar fields \cite{34,35,36,37,38}.
This model has been used to describe the structure of dark energy through the correspondence between ghost dark energy and the complex quintessence. In this respect, the authors of \cite{39} concluded that by considering the complex part of the scalar field and taking the real part of the quintessence field to be a "slow-rolling" field, a non-interacting case can not describe the real universe. Because in their calculation the $\Omega_{D}$ is greater than 1.
They also performed this test for the interacting case and assumed $\Omega_{D}=0.073$ and obtained the value of coupling parameter $b^{2}$ around $0.0849$. They showed that there is a relationship between $\Omega_{D}$ and $b^{2}$ in an interacting case.
This paper also intends to challenge the correspondence between Tsallis and Kaniadakis holographic dark energy in the Brans-Dicke cosmology concerning the complex part of the scalar field (quintessence). We will examine a dynamic model, describing the universe's acceleration in a dynamic framework that can lead to exciting results.\\
The description of the universe's accelerated expansion using dark energy has been studied in different contexts and structures. For example, the Brans–Dicke (BD) theory of gravity is an alternative to the general relativity theory.
In this theory, the gravitational constant $G$ is not constant and is replaced by the inverse of a scalar field. Here we note that the theoretical predictions for some parameters of this theory have a  significant difference with last observable data \cite{40,41,42,43}.
Holographic dark energy has been studied in a theoretical structure of Brans-Dicke (BD). It has even been argued that since holographic dark energy is a dynamic model, one should check it in dynamic frameworks such as BD\cite{44,45,46,47,48,49,50,51,52,53}.\\
These explanations motivated us to use the holographic dark energy structure from a complex quintessence point of view and within the framework of Brans–Dicke theory for the Kaniadakis and Tsallis models.
In their previous work, the authors of this article studied the Tsallis holographic dark energy under the Complex form of the Quintessence model. They compared the results with the latest observable data\cite{54}. This gives us the motivation to study further and investigate Brans-Dicke theory from a complex field by considering two important structures of holographic dark energy arising from statistical mechanics. By considering all the above explantation, we organize the article as follows.\\
In section 2, we introduce the basic structure and equations of complex quintessence (CQ).
In section 3, we describe the Tsallis holographic dark energy (THDE) in the context of Brans–Dicke theory and detail the basic equations.
Then, we create some correspondence between Tsallis holographic dark energy in the framework of Brans–Dicke theory and complex quintessence in two interacting and non-interacting cases. In that case,  we study the scalar field dynamics and analyze $ \omega- \omega '$. We will follow a similar process for Kaniadakis holographic dark energy (KHDE) in Section 4. Finally, we express the results in section 5.

\section{Complex quintessence theory (CQT)}\label{s1}
First, we consider the Friedmann-Robertson-Walker (FRW) metric that is given by the following line element \cite{38},
\begin{equation}\label{1}
	ds^{2}=-dt^{2}+a^{2}(t)\big(\frac{dr^{2}}{1-kr^{2}}+r^{2}d\Omega^{2}\big),
\end{equation}
where $a(t)$ is the scale factor and $k$ is a constant that determines the curvature of space, so that $k=0\,,-1\,,1$ for the flat and closed universe, respectively. The action of the complex quintessence model is in the following form
\begin{equation}\label{2}
	\mathcal{S}=\int d^{4}x\sqrt{-g}\big(\frac{1}{16\pi G}R+\mathcal{L}_{m}+\mathcal{L}_{\Phi}\big),
\end{equation}
where $g$ is the determinant of the metric tensor $g_{\mu\nu}$, $G$ is Newton’s constant, $R$ is  Ricci scalar and $\mathcal{L}_{m}$ is the Lagrangian density of the ordinary matter field. Also, the  Lagrangian density of complex quintessence theory (CQT)  define  as follows
\begin{equation}\label{3}
	\mathcal{L}_{\Phi}=\frac{1}{2}g^{\mu\nu}(\partial_{\mu}\Phi^{\star})(\partial_{\nu}\Phi)-V(|\Phi|),
\end{equation}
where $ \mu, \nu =$ 0, 1, 2, 3 . In the above equation, it was  assumed that the potential has only depends on the values of the complex quintessence scalar field. Now we introduce a new parameter for rewriting the complex forms of quintessence scalar field with the phase $\theta $ and amplitude $\phi$.
\begin{equation}\label{4}
	\Phi(x)=\phi(x)\exp\big(i\theta(x)\big)
\end{equation}
Another point that should be made here is that a more accurate form of equation ($\ref{4}$) can be made by definition $ \Phi(t) = \phi (t) \exp \big (i \theta (t) \big) $.  Note that, the new variables help us to reconstruct the new equations, which will connect the SNe Ia data and the quintessence potential. The Lagrangian density in terms of the newly defined variables is given by
\begin{equation}\label{5}
	\mathcal{L}_{\phi}=\frac{1}{2}g^{\mu\nu}(\partial_{\mu}\phi)(\partial_{\nu}\phi)+\frac{1}{2}\phi^{2}g^{\mu\nu}(\partial_{\mu}\theta)(\partial_{\nu}\theta)-V(\phi)\,.
\end{equation}
Now, by variation of action $(\ref{2})$ with respect to the metric and scalar field and using the FRW metric $(\ref{1})$, we obtain the field equations and the equation of motion for the scalar field as follows, respectively
\begin{equation}\label{6}
	H^{2}\equiv(\frac{\dot{a}}{a})^{2}=\frac{8\pi G}{3}\rho-\frac{k}{a^{2}}=\frac{8\pi G}{3}\left[\rho_{m}+\frac{1}{2}(\dot{\phi}^{2}+\phi^{2}\dot{\theta}^{2})+V(\phi)\right]-\frac{k}{a^{2}},
\end{equation}

\begin{equation}\label{7}
	(\frac{\ddot{a}}{a})^{2}=-\frac{4\pi G}{3}(\rho+3p)=-\frac{8\pi G}{3}\left[\frac{1}{2}\rho_{m}+(\dot{\phi}^{2}+\phi^{2}\dot{\theta}^{2})-V(\phi)\right]\big),
\end{equation}

\begin{equation}\label{8}
	\ddot{\phi}+3H\dot{\phi}-\dot{\theta}^{2}\phi+V'(\phi)=0,
\end{equation}
and
\begin{equation}\label{9}
	\ddot{\theta}+(2\frac{\dot{\phi}}{\phi}+3H)\dot{\theta}=0\,.
\end{equation}

Here $H$ is the Hubble parameter $ p $ and $ \rho $ are the pressure and the energy density of the ordinary matter and $ dot $ and prime denote a derivative with respect to time and $ \Phi $, respectively. These equations determine the main governing equations of the evolution of the universe, so that equations ($\ref{6}$) and ($\ref{7}$) represent the Friedman equations. Due to the evolution of a complex form of the scalar field, the energy density $ \rho_{\Phi} $ and the pressure $p _{\Phi} $,  is obtained by

\begin{equation}\label{10}
	\rho_{\Phi}=\frac{1}{2}\left[\dot{\phi}^{2}+\phi^{2}\dot{\theta}^{2}\right]+V(\phi),
\end{equation}
and
\begin{equation}\label{11}
	p_{\Phi}=\frac{1}{2}(\dot{\phi}^{2}+\phi^{2}\dot{\theta}^{2})-V(\phi)\,.
\end{equation}
Solving equation ($\ref{9}$) yields a solution for angular velocity in the following form
\begin{equation}\label{12}
	\dot{\theta}=\frac{\omega}{a^{3}\phi^{2}}.
\end{equation}

The parameter $ \omega $ is a constant that is determined by the initial conditions on $ \dot {\theta}$. As mentioned above,  all equations ($\ref{6}-\ref{9}$) can be rewritten according to the scalar field $\phi$. In the following sections, we will first state the main equations of the holographic dark energy scenario in the BD framework for the two specific models; Tsallis and Kaniadakis. Then we proceed with the correspondence between CQT and THDE as well as CQT and KHDE in terms of two cases viz non-interacting and interacting. We will compare each case to the latest observable data as well as with respect to each other.

\section{Tesallis holography dark energy  in the Brans-Dicke cosmology}\label{s2}

Recently, using the generalized entropy and considering the Hubble horizon as the IR cutoff, a new version of holographic dark energy that is called Tsallis holographic dark energy (THDE) has been introduced. In this article, we also want to study a new application of this model. Generally we know that  the generalized Tsallis entropy-area is independent of gravity. On the other hand, the energy density for this structure  concerning the Hubble horizon $L=H^{-1}$ is expressed by the following equation \cite{55,56,57,58,59,60,61},
\begin{equation}\label{13}
\rho_{D}=B\phi^{2\delta}H^{4-2\delta}
\end{equation}
where $B$ is unknown parameter and $\phi^{2}=\omega/(2\pi G_{eff})$ where $G_{eff}$ is the effective gravitational constant. Two important points arises here from the reduced $G_{eff}$ to $G$ and  $\delta=1$. In the first case, the THDE energy density is restored in standard cosmological studies and in the second case, the equation ($\ref{13}$) delivers the standard density of the holographic dark energy in BD gravity. Here, the fractional energy densities of the dark energy, dark matter, curvature and scalar field are expressed as
\begin{equation}\label{14}
\begin{split}
&\Omega_{D}=\frac{\rho_{D}}{\rho_{cr}}=\frac{4B\omega}{3}\phi^{2\delta-2}H^{2\delta-2},\\
&\Omega_{m}=\frac{\rho_{m}}{\rho_{cr}}=\frac{4\omega \rho_{m} }{3\phi^{2}H^{2}},\\
&\Omega_{k}=\frac{k}{H^{2}a^{2}},\\
&\Omega_{\phi}=2n(\frac{n\omega}{3}-1),
\end{split}
\end{equation}
respectively. If we assume that there is no energy exchange between dark matter (DM)  and dark energy (DE), i.e., that there is no interaction, so we will have,
\begin{equation}\label{15}
\dot{\rho}_{D}+3H\rho_{D}(1+\omega_{D})=0,
\end{equation}
and
\begin{equation}\label{16}
\dot{\rho}_{m}+3H\rho_{m}=0
\end{equation}
 where we have assumed a pressureless dust matter  for DM and $\omega_{D}=\frac{p_{D}}{\rho_{D}}$ is the equation of state (EoS) parameter of dark energy that is obtained as
\begin{equation}\label{17}
\omega_{D}=-1-\frac{2\delta n}{3}+(\delta-2)\times\frac{3(\Omega_{D}-1)-\Omega_{k}+2n(\delta \Omega_{D}+n\omega+\frac{2\omega n^{2}}{3}-2n-\Omega_{k}-4)}{3(\delta-2)\Omega_{D}-2n^{2}\omega+6n+3}.
\end{equation}
Unlike the previous case, if there is an interaction between different parts of the universe, the energy exchange can take place. In this case, we will have,
\begin{equation}\label{18}
\dot{\rho}_{D}+3H\rho_{D}(1+\omega_{D})=-\mathcal{Q},
\end{equation}
and
\begin{equation}\label{19}
\dot{\rho}_{m}+3H\rho_{m}=\mathcal{Q}
\end{equation}
In the above equation, $\mathcal{Q}$ is an interaction term which is expressed as $\mathcal{Q}=3b^{2}qH(\rho_{m}+\rho_{D})$, where $q=-1-\frac{\dot{H}}{H^{2}}$ is the deceleration parameter and $b^{2}$ is the coupling constant parameter. With these definitions, the equation of state is mentioned in the interacting case is found as follows
\begin{equation}\label{20}
\begin{split}
&\omega_{D}=-1-\frac{2\delta n}{3}+\frac{(2n-\frac{2\omega n^{2}}{3}+1)b^{2}}{\Omega_{D}}+\frac{(6n-2\omega n^{2}+3)b^{2}+2(\delta-2)\Omega_{D}}{6\Omega_{D}}\\
&\times\left[3\left(\Omega_{D}-(1+\Omega_{k})(1+b^{2})\right)-2\Omega_{k}(n-1)+2n\left(\delta\Omega_{D}+\frac{2\omega n^{2}}{3}+(n\omega-3)(b^{2}+1)-2n-1\right)\right]\\
&\times\left[2(\delta-2)\Omega_{D}-\frac{4\omega n^{2}}{3}+(3b^{2}+2)(2n+1)+b^{2}(3\Omega_{k}-2n^{2}\omega)\right]^{-1}
\end{split}
\end{equation}
Now we will take a similar process to the previous model and examine the correspondence between CQT and KHDE in the BD cosmology. We also compare the results with the previous model. You can see Ref. \cite{62,63,64,65,66,67,68,69,70,71} for further study about KHDE.

\subsection{Non-interacting case}
Here we first establish a correspondence between THDE in the context of BD cosmology with CQT for the non-interacting case. So, by combining equations $(\ref{10})$, $(\ref{12})$ and $(\ref{13})$, we get
\begin{equation}\label{21}
	\rho_{D}=\frac{1}{2}\big(\dot{\phi}^{2}+\frac{\omega^{2}}{a^{6}\phi^{2}}\big)+V(\phi)=B\phi^{2}H^{4-2\delta}
\end{equation}
for simplicity, we define
\begin{equation}\label{22}
	\mathcal{W}=\frac{1}{2}\big(\dot{\phi}^{2}+\frac{\omega^{2}}{a^{6}\phi^{2}}\big).
\end{equation}
combining the equations \ref{21} and \ref{22}, one can obtain
\begin{equation}\label{23}
	V(\phi)=B\phi^{2}H^{4-2\delta}-\mathcal{W}.
\end{equation}
To establish a correspondence between the energy density of CQT and THDE in BD cosmology, we consider a relation as $\omega_{\Phi}\equiv\frac{p_{\Phi}}{\rho_{\Phi}}=\omega_{D}$. So, using equations $(\ref{10})$, $(\ref{11})$ and $(\ref{17})$, we get the following expression,
\begin{equation}\label{24}
	\begin{split}
		&\frac{\mathcal{W}-V(\phi)}{\mathcal{W}+V(\phi)}=\omega_{D}=-1-\frac{2\delta n}{3}+(\delta-2)\\
		&\times\frac{3(\Omega_{D}-1)-\Omega_{k}+2n(\delta \Omega_{D}+n\omega+\frac{2\omega n^{2}}{3}-2n-\Omega_{k}-4)}{3(\delta-2)\Omega_{D}-2n^{2}\omega+6n+3}.
	\end{split}
\end{equation}
By solving equation $(\ref{24})$, one can easily calculate the potential in the following form.
\begin{equation}\label{25}
	\begin{split}
		V(\phi)=\frac{\mathcal{W}\big(-3\Omega_{k}(1+2n)(-2+\delta)+(4n(-1+\delta)+3\delta)(-3+2n(-3+n\omega))-9(-2+\delta)\Omega_{D}\big)}{3\Omega_{k}(1+2n)(-2+\delta)-(3(-2+\delta)+4n(-1+\delta))(-3+2n(-3+n\omega))-9(-2+\delta)\Omega_{D}}
	\end{split}
\end{equation}
Using the equation $(\ref{23})$ together eq. $(\ref{25})$, we have
\begin{equation}\label{26}
	\begin{split}
		&\frac{\mathcal{W}\big(-3\Omega_{k}(1+2n)(-2+\delta)+(4n(-1+\delta)+3\delta)(-3+2n(-3+n\omega))-9(-2+\delta)\Omega_{D}\big)}{3\Omega_{k}(1+2n)(-2+\delta)-(3(-2+\delta)+4n(-1+\delta))(-3+2n(-3+n\omega))-9(-2+\delta)\Omega_{D}}\\
		&=B\phi^{2}H^{4-2\delta}-\mathcal{W}.
	\end{split}
\end{equation}
Assuming a flat FRW universe ($k=0$), the Hubble parameter can be calculated as
\begin{equation}
	H=\bigg(\frac{\mathcal{X}}{\mathcal{Y}}\bigg)^{\frac{1}{4-2\delta}}
\end{equation}

where we have defined
\begin{equation}\label{27}
	\begin{split}
		&\mathcal{X}=6\mathcal{W}(-3+2n(-3+n\omega)-3(-2+\delta)\Omega_{D})\\
		&\mathcal{Y}=B\phi^{2}(3\Omega_{k}(1+2n)(-2+\delta)-(3(-2+\delta)+4n(-1+\delta))(-3+2n(-3+n\omega))-9(-2+\delta)\Omega_{D})\\
		\nonumber
	\end{split}
\end{equation}
Now we use a powerful tool to distinguish the different dark energy models, i.e., $\omega-\omega'$ for the above mentioned model in the non-interacting shame. We will analyze the results by plotting some figures. As stated earlier, $\omega$ is the equation of state, and $\omega'$ is a derivative from $\omega$ versus $\ln a$. To study $\omega-\omega'$ , for the non-interacting case, we take a time derivative of equation ($\ref{13}$) and consider $\dot{\Omega}_{D}=H\Omega'_{D}$, so that leads to the following relation
\begin{equation}\label{28}
	\begin{split}
		\Omega'_{D}=2(1-\delta)\Omega_{D}\big(\frac{\dot{H}}{H^{2}}+n\big)
	\end{split}
\end{equation}
where,
\begin{equation}\label{29}
	\begin{split}
		&\frac{\dot{H}}{H^{2}}=\bigg(3(\Omega_{D}-1)-\Omega_{k}+2n(\delta\Omega_{D}+\frac{2\omega n^{2}}{3}+n\omega-2n-\Omega_{k}-4)\bigg)\\
		&\times\bigg[2(\delta-2)\Omega_{D}-\frac{4\omega n^{2}}{3}+4n+2\bigg]^{-1}
	\end{split}
\end{equation}
Hence, by combining equations ($\ref{17}$), ($\ref{28}$) and ($\ref{29}$), we get
\begin{equation}\label{30}
	\omega'=\frac{(-1+\delta)\omega_{D}(3+4n+3\omega_{D})\left(\mathcal{Z}+[9-6n(-3+n\omega)]\omega_{D}\right)}{(-2+\delta)\mathcal{Z}}
\end{equation}

where we have set $$\mathcal{Z}\equiv3\Omega_{k}(1+2n)(-2+\delta)-(3+4n)(-1+\delta)[-3+2n(-3+n\omega)]$$
In figure 1 we have plotted $ \omega_{D}-\omega_{D}'$ for various values of the parameter $\delta$. As the figure shows,
\begin{figure}[h!]
	\begin{center}
		\subfigure[]{
			\includegraphics[height=5cm,width=5cm]{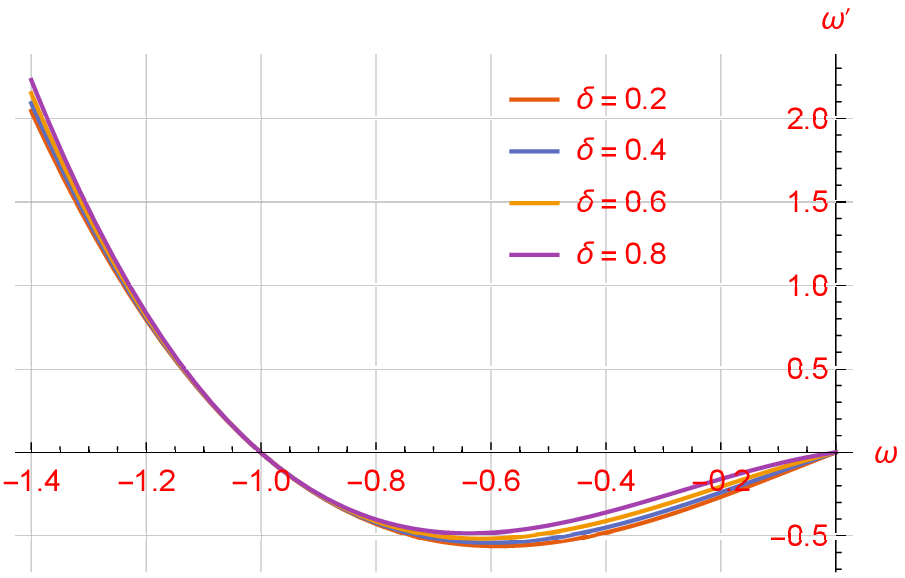}
			\label{1a}}
		\subfigure[]{
			\includegraphics[height=5cm,width=5cm]{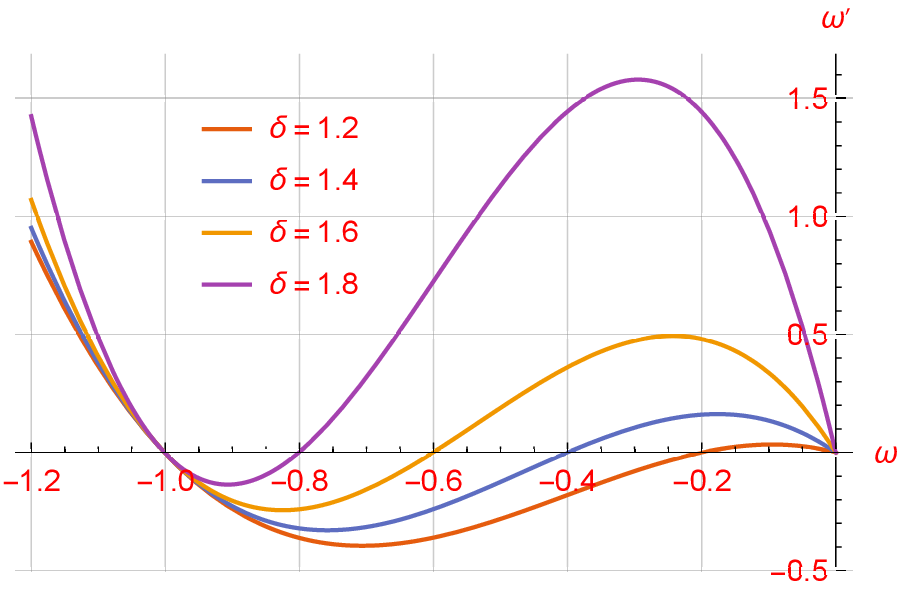}
			\label{1b}}
		\subfigure[]{
			\includegraphics[height=5cm,width=5cm]{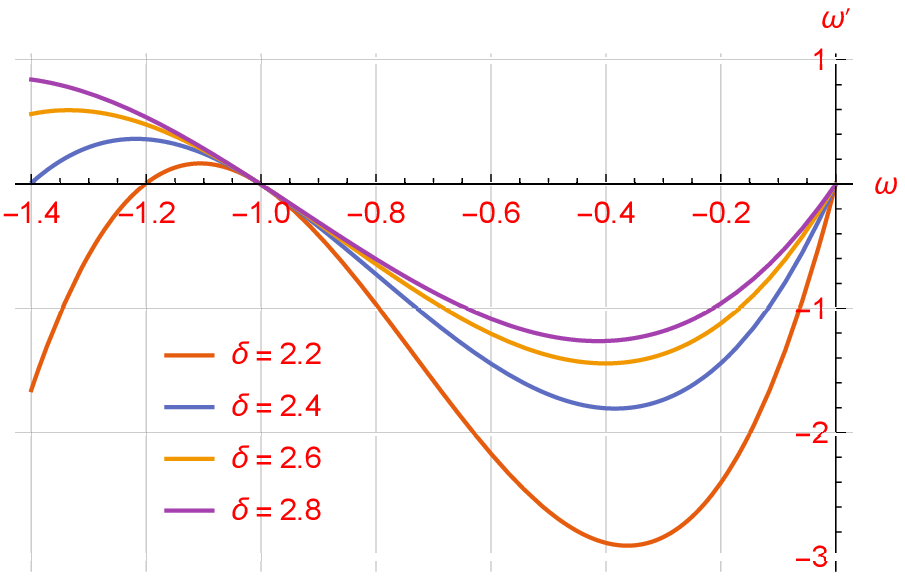}
			\label{1c}}
		\caption{\small{$\omega_{D}-\omega_{D}'$ analyze for the non-interacting case and concerning constant mentioned parameters $\omega=1000$, $n=0.001$, $k=0$ and various values of $\delta$}}
		\label{fig.1}
	\end{center}
\end{figure}
In figure ($\ref{fig.1}$) we have plotted $\omega_{D}-\omega_{D}'$ for various values of the parameter $\delta$. As the figure shows,
for $0<\delta<1$ and $\omega_{D}=-0.6$, $\omega_{D}'$ has a  minimum. In addition, as it is clear form figure $\ref{1a}$, $\omega_{D}'$ tend the zero for $\omega=0, -1$ and it takes negative values in the range of $-1<\omega_{D}<0$.
When $1<\delta<2$, $\omega_{D}'$ represents a different behavior; it has a maximum and a minimum for some values of $\omega_{D}$. Similar to the previous example, $\omega_{D}'$ has zero values when $\omega_{D}=0$ or $\omega_{D}=-1$.
Here one can say that,  when the parameter $\delta$ takes values between 2 and 3, the results are similar to the first case. Changing other free parameters in the allowable range, similar results can be obtained too. In this case, the variation of figure per parameter $\delta$ seems more tangible.
The values of the free parameter are specified in \cite{27} and these results are in agreement with the case reviewed by the authors in \cite{39}.
For example, in the case of interacting ghost dark energy in complex
quintessence theory, the free parameters were not observed, and one can obtain the change of the figure only for different values of $\omega_{D}$.
In what follows, we will review the obtained results for an interacting case and compare them with the results of this section.
We will analyze and evaluate the results by plotting some figures and setting free parameters.
In addition, we aim to study the effects of a complex part of the quintessence field. So we limit our calculations to the impact of the slow-rolling field and study the effect of free parameters. By considering the complex part of the scalar field and taking the real part of the quintessence field to be a "slow-rolling" field, we determine the relationship between these parameters and $\Omega_{D}$ which plays a significant role in determining the real universe.
For this purpose,  according to equations ($\ref{10}$), ($\ref{11}$) and ($\ref{22}$), we will have
\begin{equation}\label{31}
	\frac{1}{2}\big(\dot{\phi}^{2}+\frac{\omega^{2}}{a^{6}\phi^{2}}\big)=-\frac{B H^{4-2\delta}\phi^{2}\mathcal{Z}}{9-6n(-3+n\omega)+9(-2+\delta)\Omega_{D}}
\end{equation}
Given that in this paper we have assumed slow-roll, we can ignore the $\dot{\phi}^{2}$. Then, with straightforward calculations, we get two values for $\phi$ as follows
\begin{equation}\label{32}
	\phi=\bigg(-\frac{a^{6}}{2}\mp\bigg[\frac{1}{4}a^{12}		+\frac{3\omega H^{2\delta-4}}{B\mathcal{Z}}(-3+2n(-3+n\omega)-3(-2+\delta)\Omega_{D})\bigg]^{\frac{1}{2}}
	\bigg)^{\frac{1}{2}}
\end{equation}
Here, we consider the positive sign of the above solution. Since $\dot{\phi}=H\frac{d\phi}{d\ln a}$ and taking into account the fact that in general $H$ is not vanishing parameter, so we must have $\frac{d\phi}{d\ln a}\approx0$. Therefore, according to the equations ($\ref{28}$), ($\ref{29}$) and ($\ref{32}$), we obtain the following result
\begin{equation}\label{32}
	\begin{split}
		&0\approx\frac{d\phi}{d\ln a}=\bigg(9H^{-2+2\delta}(-2+\delta)(-1+\delta)\Omega_{D}\big(-3\Omega_{k}(1+2n)+(3+4n)(-3-6n+2n^{2}\omega+3\Omega_{D})\big)\bigg)\\
		&\bigg/\bigg\{\sqrt{2B\mathcal{Z}}\big(3+6n-2n^{2}\omega+3(-2+\delta)\Omega_{D}\big)\bigg[a^{12}BH^{4}\mathcal{Z}+12H^{2\delta}\omega(-3+2n(-3+n\omega)-3(-2+\delta)\Omega_{D})\bigg]^{1/2}\\
		&\times\bigg(-\frac{a^{6}}{2}\mp\bigg[\frac{1}{4}a^{12}		+\frac{3\omega H^{2\delta-4}}{B\mathcal{Z}}(-3+2n(-3+n\omega)-3(-2+\delta)\Omega_{D})\bigg]^{\frac{1}{2}}
		\bigg)^{\frac{1}{2}}\bigg\}
	\end{split}
\end{equation}
Permissible values $\Omega_{D}$ can be calculated by setting free parameters.
Of course, given the values we have for free parameters, it has several answers, some of which are acceptable and some of which are not, in the sense that $\Omega_{D}$ should be less than 1.
In this article, we will consider specific conditions in each step to assess the particular points that we will explain in detail.
\begin{equation}\label{32}
	\begin{split}
		9(-2+\delta)(-1+\delta)\Omega_{D}\big(-3\Omega_{k}(1+2n)+(3+4n)(-3-6n+2n^{2}\omega+3\Omega_{D})\big)=0
	\end{split}
\end{equation}
It is worth noting that, assuming the free parameters mentioned in the text, the computational values for $\Omega_{D}$ can take values not only less than $1$ but also can be greater than unity. This is an essential point that we will explain in detail. From equation ($\ref{32}$), a correlation is observed between the free parameters for the THDE in BD cosmology and the parameter $\Omega_{D}$.
By setting free parameters, we can obtain the allowable values of parameter $\Omega_{D}$.
In exchange for these free parameters, the values obtained for $\Omega_{D}$ may be in the two ranges, i.e., smaller and greater than 1, which is also the case for the following example.
The answer is acceptable for those solutions that $\Omega_{D}<1$, but for those solutions that lead to $\Omega_{D}>1$, the non-interacting case can not describe the real universe.
Therefore, we will need to analyze the interacting case.
In this model, you will see that there will be a close relationship between the coupling parameter $b^{2}$ with  $\Omega_{D}$ and $\delta$, which we will explain in detail.

\subsection{Interacting case}
Here, we consider the previous process for the interacting case to perform the calculations and compare the results with the non-interacting case. Hence for simplicity, according to equation ($\ref{20}$), we have.
\begin{equation}\label{32}
\begin{split}
&\mathcal{O}=-\omega_{D}=-\bigg(-1-\frac{2\delta n}{3}+\frac{(2n-\frac{2\omega n^{2}}{3}+1)b^{2}}{\Omega_{D}}+\frac{(6n-2\omega n^{2}+3)b^{2}+2(\delta-2)\Omega_{D}}{6\Omega_{D}}\\
&\times\big\{3(\Omega_{D}-(1+\Omega_{k})(1+b^{2}))-2\Omega_{k}(n-1)+2n(\delta\Omega_{D}+\frac{2\omega n^{2}}{3}+(n\omega-3)(b^{2}+1)-2n-1)\big\}\\
&\times\big[2(\delta-2)\Omega_{D}-\frac{4\omega n^{2}}{3}+(3b^{2}+2)(2n+1)+b^{2}(3\Omega_{k}-2n^{2}\omega)\big]^{-1}\bigg).
\end{split}
\end{equation}
In the non-interacting case, we also mentioned that since the aim is to establish a correspondence between THDE and CQT in BD cosmology, then we assume that $\omega_{\Phi}=\omega_{D}$, hence we have
\begin{equation}\label{33}
\begin{split}
-\mathcal{O}=\frac{\mathcal{W}-V(\phi)}{\mathcal{W}+V(\phi)}.
\end{split}
\end{equation}
also one can obtain,
\begin{equation}\label{34}
\begin{split}
V(\phi)=-\frac{\mathcal{O}+1}{\mathcal{O}-1}\mathcal{W}.
\end{split}
\end{equation}
Here, it will be calculated by two equations ($\ref{23}$) and ($\ref{34}$),
\begin{equation}\label{35}
\begin{split}
-\frac{\mathcal{O}+1}{\mathcal{O}-1}\mathcal{W}=B\phi^{2}H^{4-2\delta}-\mathcal{W}.
\end{split}
\end{equation}
So we can calculate,
\begin{equation}\label{36}
\begin{split}
H=(-\frac{2\mathcal{W}}{B(-1+\mathcal{O})\phi^{2}})^{\frac{1}{4-2\delta}}=(-\frac{2\mathcal{W}}{B(-1-\omega_{D})\phi^{2}})^{\frac{1}{4-2\delta}}.
\end{split}
\end{equation}
To be self-consistent, the $\mathcal{O}$ must be less than 1. In addition, according to equation ($\ref{7}$) and to have accelerated expansion of the universe, there must be a constraint as $\rho_{m}<2V(\phi)-(\dot{\phi}^{2}+\phi^{2}\dot{\theta}^{2}))=2V(\phi)-4\mathcal{W}$.
As a result, by combining with equation ($\ref{23}$), one can rewrite as $\rho_{m}<2V(\phi)-4\mathcal{W}=6V(\phi)-4B\phi^{2}H^{4-2\delta}$.
Hence, according to the above mentioned points, we can conclude that $V(\phi)>2/3 B\phi^{2}H^{4-2\delta}$.
Also, using the above mentioned relations and equation $\ref{23}$), it is possible to determine a restriction for potential in the following form,
\begin{equation}\label{37}
\begin{split}
2/3 B\phi^{2}H^{4-2\delta}<V(\phi)<4B\phi^{2}H^{4-2\delta}.
\end{split}
\end{equation}
Now we will examine  $\omega_{D}-\omega_{D}'$ for the interacting case and compare the similarities and differences with the non-interacting case. Of course, it should be noted that we will repeat the computational process for KHDE in the following sections. So we will have a time derivative of equation $(\ref{13})$ and combine with relation $\dot{\Omega}_{D}=H\Omega'_{D}$. As previous, prime represents a derivative concerning $x = \ln a$, so for this case we obtain the following equation
\begin{equation*}\label{37}
\begin{split}
\Omega'_{D}=2(1-\delta)\Omega_{D}(\frac{\dot{H}}{H^{2}}+n)
\end{split}
\end{equation*}
where
\begin{equation*}\label{37}
\begin{split}
&\frac{\dot{H}}{H^{2}}=\bigg(3\Omega_{D}-3(1+\Omega_{k})(1+b^{2})-2\Omega_{k}(n-1)+2n(\delta\Omega_{D}+\frac{2\omega n^{2}}{3}+(n\omega-3)(b^{2}+1)-2n-1)\bigg)\\&\times\bigg[2(\delta-2)\Omega_{D}-\frac{4n^{2}\omega}{3}+(3b^{2}+2)(2n+1)+b^{2}\big(3\Omega_{k}-2n^{2}\omega\big)\bigg]^{-1}.
\end{split}
\end{equation*}
Using the equations ($\ref{13}$), ($\ref{14}$) and ($\ref{20}$), we get
\begin{equation*}
	\begin{split}
&\mathcal{I}=\big(-3+2n(-3+n\omega)\big)\bigg(6\delta+4n(-2+3\delta)+3b^{2}(7+2(-1+n)\delta)+6(2+3b^{2})\omega_{D}\\
&+\bigg[\bigg(9b^{2}(1-2(1+n)\delta)^{2}+12b^{2}(-1+2(1+n)\delta)(-3(-4+\delta)+2n(2+\delta))\\
&+4(3\delta+n(-4+6\delta))^{2}+12\omega_{D}(6b^{2}(3+4n)(1+\delta)+9b^{2}(-1+2(1+n)\delta)+4(3\delta+n(-4+6\delta))\\
&+3(2+3b)^{2}\omega_{D})\bigg)\bigg]^{\frac{1}{2}}\bigg/\big(12(-2+\delta)(3+2n\delta+6\omega_{D})\big)\bigg)\,,
\end{split}
\end{equation*}

\begin{equation*}
	\begin{split}
&\mathcal{J}=2(-1+\delta)\Omega_{D}\bigg\{n-\bigg(4n^{2}(-3+(3+n)\omega)+3b^{2}(-3+2n(-3+n\omega))\\
&+9(-1+\Omega_{D})+6n(-4+\delta\Omega_{D})\bigg/\bigg[6-4n(-3+n\omega)+3b^{2}(3+6n-4n^{2}\omega)+6(-2+\delta)\Omega_{D}\bigg]\bigg)\bigg\}\\
	\end{split}
\end{equation*}

\begin{equation*}
	\begin{split}
&\mathcal{E}_{1}=2b^{2}\mathcal{J}(-3+2n(-3+n\omega))\bigg/6\mathcal{I}^{2}\\
&\mathcal{E}_{2}=+3\mathcal{J}(3+2n\delta)\mathcal{I}\times(b^{2}(-3+2n(-3+n\omega))-2(-2+\delta)\mathcal{I})\bigg/6\mathcal{I}^{2}(2+3b^{2})(-3+2n(-3+n\omega))\\
&-6(-2+\delta)\mathcal{I}\\
&\mathcal{E}_{3}=-\mathcal{J}(b^{2}(-3+2n(-3+n\omega))-2(-2+\delta)\mathcal{I})((3+3b^{2}+2n)(-3+2n(-3+n\omega))\\&+3(3+2n\delta)\mathcal{I})\bigg/(2+3b^{2})(-3+2n(-3+n\omega))-6(-2+\delta)\mathcal{I}
	\end{split}
\end{equation*}
\begin{equation}\label{38}
\begin{split}
&\mathcal{E}_{4}=-6\mathcal{J}(-2+\delta)\mathcal{I}(b^{2}(3+6n-2n^{2}\omega)+2(-2+\delta)\mathcal{I})((3+3b^{2}+2n)\\
&\times(-3+2n(-3+n\omega))+3(3+2n\delta)\mathcal{I})\bigg/6\mathcal{I}^{2}((2+3b^{2})(-3+2n(-3+n\omega))-6(-2+\delta)\mathcal{I})^{2}\\
&\mathcal{E}_{5}=+2c(-2+\delta)\mathcal{I}((3+3b^{2}+2n)(-3+2n(-3+n\omega))+3(3+2n\delta)\mathcal{I})\\
&\bigg/-6\mathcal{I}^{2}(2+3b^{2})(-3+2n(-3+n\omega))+6(-2+\delta)\mathcal{I}\\
&\omega'=\mathcal{E}_{1}+\mathcal{E}_{2}+\mathcal{E}_{3}+\mathcal{E}_{4}+\mathcal{E}_{5}
\end{split}
\end{equation}
\begin{figure}[h!]
 \begin{center}
 \subfigure[]{
 \includegraphics[height=5cm,width=5cm]{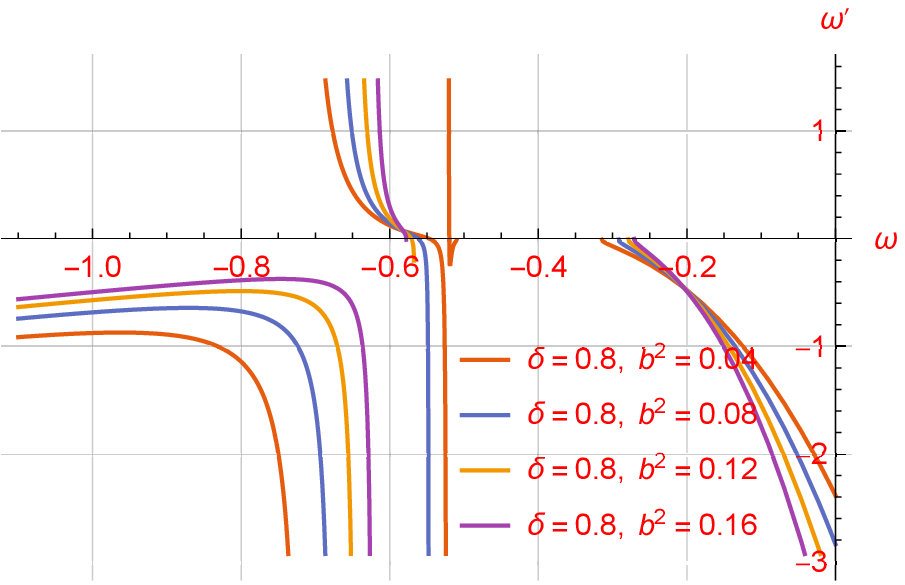}
 \label{2a}}
 \subfigure[]{
 \includegraphics[height=5cm,width=5cm]{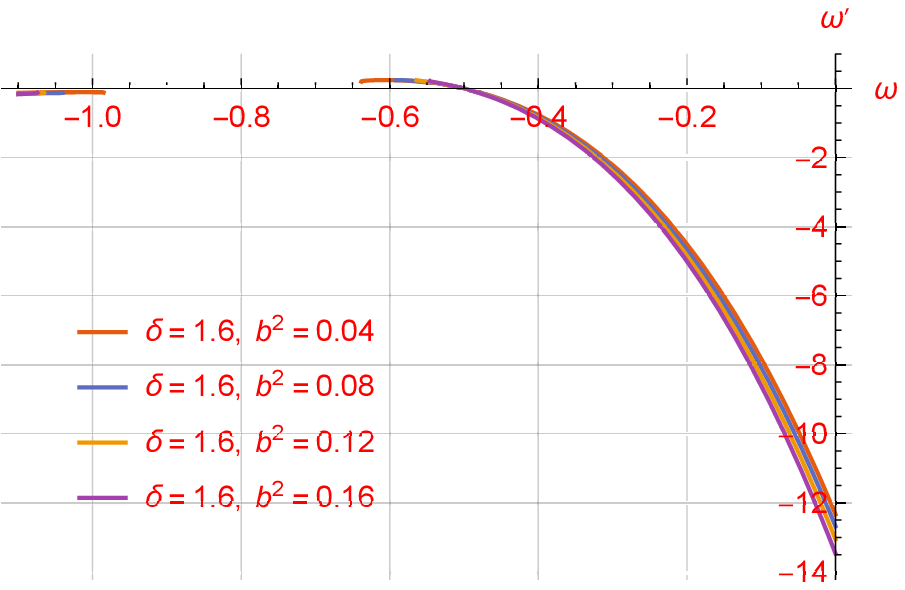}
 \label{2b}}
 \subfigure[]{
 \includegraphics[height=5cm,width=5cm]{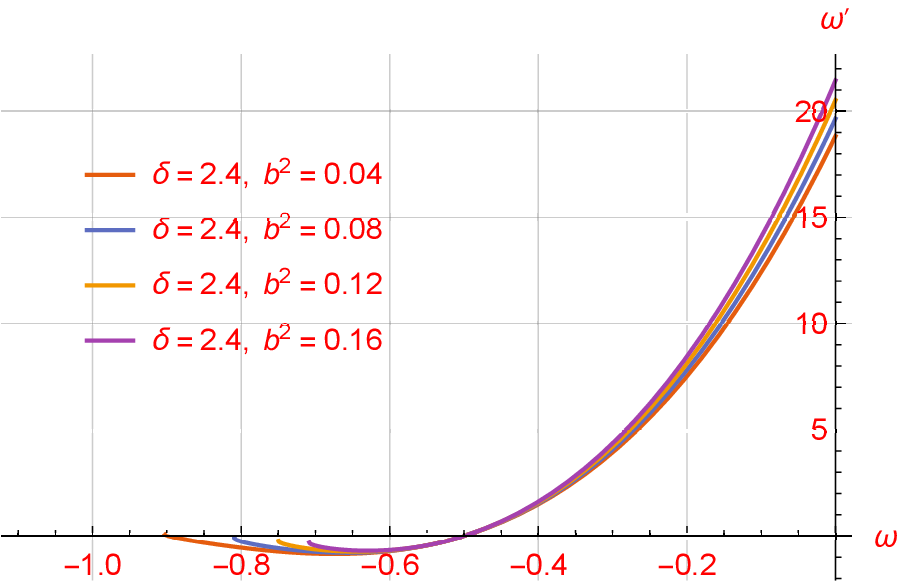}
 \label{2c}}
  \caption{\small{$\omega_{D}-\omega_{D}'$ analyze for the non-interacting case and concerning constant mentioned parameters $\omega=1000$, $n=0.001$, $k=0$ and various values of $\delta$ and $b^{2}$}}
 \label{2}
 \end{center}
 \end{figure}
In figure ($\ref{2}$) we analyze $\omega_{D}-\omega_{D}'$  for the interacting case according to the free parameters  which is expressed by  $b^{2}$ and $\delta$.
Also,  here we note that the figures are  changed significantly  by two parameters $b^{2}$ and $\delta$ .
When the parameter $\delta$ is between 0 and 1, unlike the non-interacting case, it has an overlap in $-0.6<\omega_{D}<-0.4$ and $\omega_{D}>-0.3$, it also has a maximum in $\omega_{D}<-0.6$.  Also in  range between $-0.6 <\omega_{D}< -0.2$, the parameter $\omega_{D}'$ tends to zero.
 So, this figure has both negative and positive states in the range between 0 and -1. Also, in points $\omega_{D}>-0.3$, the figure is narrower; their values  are close to each other for free parameters.

Also, the overlap and positive region of the figure is in the range ($-0.7 <\omega_{D}< -0.5$), which is different from the non-interacting case and is somewhat close to the results of  \cite{39}.
When the parameter $\delta$ is between 1 and 2, the figures obtained up to the range $\omega_{D}> -0.5$ are negative, and still, their distance is very close and narrower for different values of $b^{2}$ and $\delta$. It has overlap and a maximum in the range $\omega_{D}< -0.5$, also for $\omega_{D}= -0.5$, $\omega_{D}'$ tends to zero, which is different from the non-interacting state. Finally, when $\delta$ experiences between 2 and 3, the opposite results are obtained with the figure $\ref{2b}$ (the parameter $b^{2}$ in all three modes have similar values). However, figure $\ref{2b}$ and figure $\ref{2c}$ are not much closer to figure $\ref{2a}$. We see kind of adaptation in figure $\ref{2b}$ and figure $\ref{2c}$ with different sign.
By changing other free parameters in the allowable range, similar results are obtained. So,  in this case the figure changes per parameter $b^{2}$ and $\delta$ seem more tangible.
In the next section, we will proceed to the calculations for KHDE in BD cosmology for both interacting and non-interacting cases and compare them with respect to  each other.

As in the previous section, we intend to examine the effects of the complex part of the quintessence field for the interacting case. As mentioned earlier, by considering the complex part of the scalar field and taking the real part of the quintessence field to be a "slow-rolling" field. In the end, we will assume the results for a flat universe, so for this interacting case, like the previous part and slow-rolling conditions, we will have,

\begin{equation}\label{39}
\begin{split}
&\frac{\omega}{\phi^{2}+a^{6}}=B\phi^{2}H^{4-2\delta}\bigg(-\frac{2\delta n}{3}+\frac{(2n-\frac{2\omega n^{2}}{3}+1)b^{2}}{\Omega_{D}}+\frac{(6n-2\omega n^{2}+3)b^{2}+2(\delta-2)\Omega_{D}}{6\Omega_{D}}\\
&\times\big\{3(\Omega_{D}-(1+\Omega_{k})(1+b^{2}))-2\Omega_{k}(n-1)+2n(\delta\Omega_{D}+\frac{2\omega n^{2}}{3}+(n\omega-3)(b^{2}+1)-2n-1)\big\}\bigg)\\
&\times\big[2(\delta-2)\Omega_{D}-\frac{4\omega n^{2}}{3}+(3b^{2}+2)(2n+1)+b^{2}(3\Omega_{k}-2n^{2}\omega)\big]^{-1}
\end{split}
\end{equation}
We can obtain different solutions for the scalar field $\phi$ by solving the above equation. Here, we consider only positive solutions so that we will have
\begin{equation}\label{40}
\begin{split}
&\phi=\bigg\{\bigg(a^{6}BH^{4}(b^{2}(1+3b^{2}-2n)(3+6n-2n^{2}\omega)^{2}+(-8n+6\delta+12n\delta+3b^{2}(7+2(-1+n)\delta))\\
&\times(-3+2n((-3+n\omega))\Omega_{D}-6(-2+\delta)(3+2n\delta)\Omega_{D}^{2})\mp\bigg[BH^{4}(b^{2}(1+b^{2}-2n)\\
&\times(3+6n-2n^{2}\omega)^{2}+(-8n+6\delta+12n\delta+3b^{2}(7+(2(-1+n)\delta))(-3+2n))\Omega_{D}\\
&-6(-2+\delta)(3+2n\delta)\Omega_{D}^{2})\times(-24H^{2\delta}\omega\Omega_{D}((2+3b^{2})(-3+2n(-3+n\omega))-6(-2+\delta)\\
&+a^{12}BH(b^{2}(1+3b^{2}-2n)(3+6n-2n^{2}\omega)^{2}+(21b^{2}-8n+6(1+b^{2}(-1+n)+2n)\delta)\\
&\times(-3+n\omega))\Omega_{D}-6(-2+\delta)(3+2n\delta)\Omega_{D}^{2}))\bigg]^{\frac{1}{2}}\bigg)^{\frac{1}{2}}\bigg\}\\
&\bigg/\bigg(\sqrt{2}\bigg(BH^{4}(-b^{2}(1+3b^{2}-2n)(3+6n-2n^{2}\omega)^{2}-(6\delta+4n(-2+3\delta)+3b^{2}(7+2(-1+n)\delta))\\
&\times(-3+2n(-3+n\omega))\Omega_{D}+6(-2+\delta)(3+2n\delta)\Omega_{D}^{2})\bigg)^{\frac{1}{2}}\bigg)
\end{split}
\end{equation}

As we explained in the previous section, because we can not ignore the parameter $H$, the  $\frac{d\phi}{d\ln a}=0$ must be equal to zero. So we first calculate the $\frac{d\phi}{d\ln a}$, then we set it zero. Finally, the following plot can show the result concerning the free parameters mentioned in this paper. As we explained for the non-interacting case, there is a link between the free parameters for THDE in BD cosmology and the $ \Omega_{D} $.
By setting free parameters, we can obtain the allowable values of the $ \Omega_{D} $.
Of course, by setting the free parameters as mentioned in the text and placing $ \Omega_{D}=0.73 $, we can calculate various values for the coupling parameter $b^{2}$.
In \cite{39} authors measured the value of this parameter $b^{2}$ as $0.0849$ with a similar process. Here by setting free parameters, we can determine the value of this coupling parameter $b^{2}$ too.
 Here, according to the free parameters and $ \Omega_ {D}=0.73 $ , we determine $b^{2}$ in terms of $\delta$ by plotting the figure $\ref{3}$
Of course, this solidarity should not be easily overlooked. Of course, we can point out that the parameters can be adjusted within their allowable range so that the theory can describe the real universe, i.e., $ \Omega_{D}<1 $.

\begin{figure}[h!]
 \begin{center}
 \subfigure[]{
 \includegraphics[height=5cm,width=5cm]{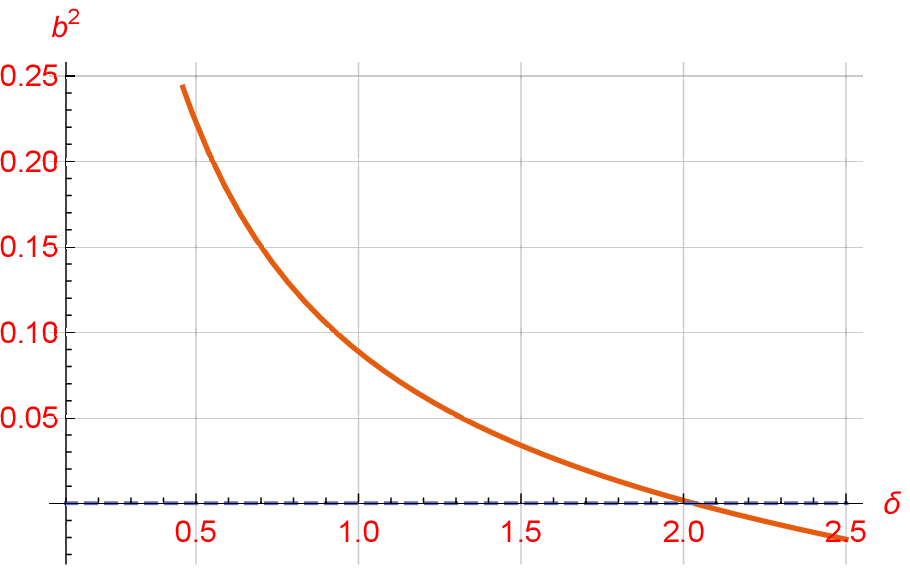}
 \label{3a}}
  \caption{\small{The plot $b^{2}$ in terms of $\delta$ concerning constant mentioned parameters $n=0.001$, $k=0$ and $\Omega_{D}=0.73$  }}
 \label{3}
 \end{center}
 \end{figure}

\section{Kaniadakis holographic dark energy(KHDE) in the BD cosmology}\label{s3}

It has been shown that the entropy of Kaniadakis is expressed as $S_{k}=\frac{1}{k}\sinh(kS_{BH})$ in which $k$ is a free parameter. Due to the thermodynamic importance of the horizon, researchers used this horizon as the IR cutoff and introduced a newer structure of holographic dark energy with stunning results. Recently, the results of KHDE in some systems have been compared with other theories of dark energy.
This new model is called KHDE, which can describe the current accelerating phase of the present universe.
Applications of this model in the non-flat universe and other IR cutoffs have been studied.
Since the nature of dark energy is still unknown, the values of free parameter $k$ of the model can be different, and no limit can be seen in these values.
However, depending on the primary presumptions, such as the IR cutoff, different selection intervals can be considered for this parameter to be more consistent with the observational data. This article will challenge specific implications of KHDE in the context of BD cosmology and compare the results with the previous model. With all these explanations according to \cite{62,63,64,65,66,67,68,69,70,71} energy density of KHDE in the BD framework and assuming that the apparent horizon in the flat space is as the IR cut-off $L=H^{-1}$, we will have.
\begin{equation}\label{41}
\rho_{D}=\frac{3c^{2}\phi^{2}H^{4}}{4\omega k}\sinh\big(\frac{2k\pi^{2}\phi^{2}}{\omega H^{2}}\big)
\end{equation}
where $k$ is an unknown constant. In this model, assuming $G_{eff}$ is reduced to $G$ and parameter $k$ be tends to zero, thus restoring the HDE energy density in standard cosmology. According to the definition of critical density such as $\rho_{cr}=\frac{3\phi^{2}H^{2}}{4\omega}$, dimensionless fractional energy densities can be defined as follows.
\begin{equation}\label{42}
\begin{split}
&\Omega_{m}=\frac{\rho_{m}}{\rho_{cr}}=\frac{4\omega\rho_{m}}{3\phi^{2}H^{2}},\\
&\Omega_{D}=\frac{\rho_{D}}{\rho_{cr}}=\frac{4\omega\rho_{D}}{3\phi^{2}H^{2}},\\
&\Omega_{\phi}=\frac{\rho_{\phi}}{\rho_{cr}}=2\big(\frac{n\omega}{3}-1\big)\,.
\end{split}
\end{equation}
To examine KHDE in BD cosmology, we consider both interacting and non-interacting cases like the previous model. If we assume no interaction between different parts of the universe, such as DM and DE, we will have relations as $\dot{\rho}_{D}+3H\rho_{D}(1+\omega_{D})=0$ and $\dot{\rho}_{m}+3H\rho_{m}=0$.
For the non-interacting case of KHDE in the BD framework, (EOS) is calculated as follows.
\begin{equation}\label{43}
\begin{split}
\omega_{D}=-1-\frac{2n}{3}-\frac{4k\pi^{2}\phi^{2}}{3\omega H^{2}}\coth\big[\frac{2kn\pi^{2}\phi^{2}}{\omega H^{2}}\big]-\frac{4}{3}\big(1-\frac{4k\pi^{2}\phi^{2}}{3\omega H^{2}}\coth\big[\frac{2k\pi^{2}\phi^{2}}{\omega H^{2}}\big]\big)\frac{\dot{H}}{H^{2}}
\end{split}
\end{equation}
where
\begin{equation}\label{44}
\begin{split}
&\frac{\dot{H}}{H^{2}}=\bigg\{-\frac{9\Omega_{m}^{0}H_{0}^{2}\phi_{0}^{2}}{4\omega\phi^{2}H^{2}}+(2\omega n^{2}-6n-3)\frac{n}{2\omega}\\
&+\frac{3nc^{2}\pi^{2}\phi^{2}}{\omega^{2}}\cosh\big(\frac{2k\pi^{2}\phi^{2}}{\omega H^{2}}\big)+\frac{3nc^{2}H^{2}}{2\omega k}\sinh\big(\frac{2k\pi^{2}\phi^{2}}{\omega H^{2}}\big)\bigg\}\\
&\times\bigg[\frac{3c^{2}\pi^{2}\phi^{2}}{\omega^{2}}\cosh\big(\frac{2k\pi^{2}\phi^{2}}{\omega H^{2}}\big)-\frac{3c^{2}H^{2}}{\omega k}\sinh\big(\frac{2k\pi^{2}\phi^{2}}{\omega H^{2}}\big)\bigg]^{-1}\,.
\end{split}
\end{equation}
If it is assumed that there are interactions between different parts, then we will have equations as $\dot{\rho}_{D}+3H\rho_{D}(1+\omega_{D})=-Q$ and $\dot{\rho}_{m}+3H\rho_{m}=Q$.
In this relation, $Q$ is an interaction terms  which is defined by $Q=H\big(\alpha\rho_{m}+\beta\rho_{D}\big)$.
The two parameters $\alpha$ and $\beta$ are coupling constants parameters. They play an important role to selecting correctly and also  matching  the values of the free parameter with the latest observational data.
It will also lead to a relationship between these free parameters and $\Omega_{D}$.
The equation of state for this mentioned model in the interacting case will also be calculated as follows.
\begin{equation}\label{45}
\begin{split}
\omega_{D}=-1+\frac{1}{3}(\alpha-\beta-2n)-\frac{4k\pi^{2}\phi^{2}}{3\omega H^{2}}\coth\big[\frac{2kn\pi^{2}\phi^{2}}{\omega H^{2}}\big]\big(n-\frac{\dot{H}}{H^{2}}\big)+\frac{k\alpha(2\omega n^{2}-2n-1)}{9c^{2}H^{2}\sinh\big(\frac{2kn\pi^{2}\phi^{2}}{\omega H^{2}}\big)}
\end{split}
\end{equation}
where
\begin{equation}\label{46}
\begin{split}
&\frac{\dot{H}}{H}=\bigg\{-\frac{3\Omega_{m}^{0}H_{0}^{2}H_{0}^{2}}{4\phi^{2}H^{2}}+(2\omega n^{2}-6n-3)\frac{n}{6}+\frac{\alpha\Omega_{m}}{4}\\
&+\frac{nc^{2}\pi^{2}\phi^{2}}{\omega}\cosh\big(\frac{2k\pi^{2}\phi^{2}}{\omega H^{2}}\big)+\frac{c^{2}H^{2}(n+\frac{\beta}{2})}{2k}\sinh\big(\frac{2k\pi^{2}\phi^{2}}{\omega H^{2}}\big)\bigg\})\\
&\times\bigg[\frac{c^{2}\pi^{2}\phi^{2}}{\omega}\cosh\big(\frac{2k\pi^{2}\phi^{2}}{\omega H^{2}}\big)-\frac{c^{2}H^{2}}{k}\sinh\big(\frac{2k\pi^{2}\phi^{2}}{\omega H^{2}}\big)+n-\frac{\omega n^{2}}{3}+\frac{1}{2}\bigg]^{-1}
\end{split}
\end{equation}
Similar the previous model, KHDE is studied in both interacting and non-interacting cases and the correspondence between energy density CQT and KHDE in BD cosmology is formed. In addition, we compare the results of this model with the latest observational data and with THDE. You can also see Ref.s \cite{62,63,64,65,66,67,68,69,70,71} for further study of mentioned model.
\subsection{Non-interacting case}
Here we first challenge the non-interacting case for KHDE and examine the mentioned correspondence; then, we investigate the interacting sample. Hence according to equations ($\ref{10}$), ($\ref{12}$) and ($\ref{41}$), we will have.
\begin{equation}\label{47}
\rho_{D}=\frac{1}{2}\big(\dot{\phi}^{2}+\frac{\omega^{2}}{a^{6}\phi^{2}}\big)=\frac{3c^{2}\phi^{2}H^{4}}{4\omega k}\sinh\big(\frac{2k\pi^{2}\phi^{2}}{\omega H^{2}}\big)\,.
\end{equation}
Here we set,
\begin{equation}\label{48}
\mathcal{M}=\frac{1}{2}\big(\dot{\phi}^{2}+\frac{\omega^{2}}{a^{6}\phi^{2}}\big).
\end{equation}
Therefore, the potential is obtained by,
\begin{equation}\label{49}
V(\phi)=\frac{3c^{2}\phi^{2}H^{4}}{4\omega k}\sinh\big(\frac{2k\pi^{2}\phi^{2}}{\omega H^{2}}\big)-\mathcal{M}.
\end{equation}
As in the previous section, we can study forming a relation as $\omega_{D}=\omega_{\Phi}$ because we are looking to establish a correspondence between the energy density of CQT and KHDE in the BD cosmology. So, by combining equations ($\ref{10}$), ($\ref{11}$) and ($\ref{43}$), we will have.
\begin{equation}\label{50}
\begin{split}
\frac{\mathcal{M}-V(\phi)}{\mathcal{M}+V(\phi)}=-1-\frac{2n}{3}-\frac{4k\pi^{2}\phi^{2}}{3\omega H^{2}}\coth\big[\frac{2kn\pi^{2}\phi^{2}}{\omega H^{2}}\big]-\frac{4}{3}\big(1-\frac{4k\pi^{2}\phi^{2}}{3\omega H^{2}}\coth\big[\frac{2k\pi^{2}\phi^{2}}{\omega H^{2}}\big]\big)\frac{\dot{H}}{H^{2}}.
\end{split}
\end{equation}
Given the two equations ($\ref{44}$) and ($\ref{50}$), the  relation for potential can be easily calculated as follows.
\begin{equation}\label{51}
\begin{split}
&V(\phi)=-\bigg\{\bigg[\mathcal{M}\bigg(36H^{4}kn\omega^{2}\phi^{2}+72H^{4}kn^{2}\omega^{2}\phi^{2}-24H^{4}kn^{3}\omega^{3}\phi^{2}+81H^{2}k\omega^{2}\phi_{0}^{2}\Omega_{m}^{0}H_{0}^{2}\\
&-162c^{2}H^{4}k\pi^{2}\omega\phi^{4}\cosh\big[\frac{2k\pi^{2}\phi^{2}}{H^{2}\omega}\big]-90c^{2}H^{4}kn\pi^{2}\omega\phi^{4}\cosh\big[\frac{2k\pi^{2}\phi^{2}}{H^{2}\omega}\big]\\
&48H^{2}k^{2}n\pi^{2}\omega\phi^{4}\coth\big[\frac{2k\pi^{2}\phi^{2}}{H^{2}\omega}\big]-96n^{2}H^{2}k^{2}\pi^{2}\omega\phi^{4}\coth\big[\frac{2k\pi^{2}\phi^{2}}{H^{2}\omega}\big]\\
&32n^{3}H^{2}k^{2}\pi^{2}\omega^{2}\phi^{4}\coth\big[\frac{2k\pi^{2}\phi^{2}}{H^{2}\omega}\big]-108k^{2}\pi^{2}\omega\phi^{2}\phi_{0}^{2}\Omega_{m}^{0}H_{0}^{2}\coth\big[\frac{2k\pi^{2}\phi^{2}}{H^{2}\omega}\big]\\
&+144c^{2}nH^{2}k^{2}\pi^{4}\phi^{6}\cosh\big[\frac{2k\pi^{2}\phi^{2}}{H^{2}\omega}\big]\coth\big[\frac{2k\pi^{2}\phi^{2}}{H^{2}\omega}\big]-108c^{2}H^{2}k^{2}\pi^{4}\phi^{6}\cosh\big[\frac{2k\pi^{2}\phi^{2}}{H^{2}\omega}\big]\coth\big[\frac{2kn\pi^{2}\phi^{2}}{H^{2}\omega}\big]\\
&+162c^{2}H^{6}\omega^{2}\phi^{2}\sinh\big[\frac{2k\pi^{2}\phi^{2}}{H^{2}\omega}\big]+108c^{2}H^{4}k\pi^{2}\omega\phi^{4}\coth\big[\frac{2kn\pi^{2}\phi^{2}}{H^{2}\omega}\big]\sinh\big[\frac{2k\pi^{2}\phi^{2}}{H^{2}\omega}\big]\bigg)\bigg]\bigg/\\
&\bigg(36H^{4}kn\omega^{2}\phi^{2}+72H^{4}kn^{2}\omega^{2}\phi^{2}-24H^{4}kn^{3}\omega^{3}\phi^{2}-90c^{2}H^{4}kn\pi^{2}\omega\phi^{4}\cosh\big[\frac{2k\pi^{2}\phi^{2}}{H^{2}\omega}\big]\\
&-48H^{2}k^{2}n\pi^{2}\omega\phi^{4}\coth\big[\frac{2k\pi^{2}\phi^{2}}{H^{2}\omega}\big]-96n^{2}H^{4}k^{2}\pi^{2}\omega\phi^{4}\coth\big[\frac{2k\pi^{2}\phi^{2}}{H^{2}\omega}\big]\\
&+32H^{2}k^{2}n^{3}\pi^{2}\omega^{2}\phi^{4}\coth\big[\frac{2k\pi^{2}\phi^{2}}{H^{2}\omega}\big]-108k^{2}\pi^{2}\omega\phi^{2}\phi_{0}^{2}\Omega_{m}^{0}H_{0}^{2}\coth\big[\frac{2k\pi^{2}\phi^{2}}{H^{2}\omega}\big]\\
&+144c^{2}H^{2}k^{2}n\pi^{4}\phi^{6}\cosh\big[\frac{2k\pi^{2}\phi^{2}}{H^{2}\omega}\big]\coth\big[\frac{2k\pi^{2}\phi^{2}}{H^{2}\omega}\big]\\
&-108c^{2}H^{2}k^{2}\pi^{4}\phi^{6}\cosh\big[\frac{2k\pi^{2}\phi^{2}}{H^{2}\omega}\big]\coth\big[\frac{2k\pi^{2}\phi^{2}}{H^{2}\omega}\big]\\
&+108c^{2}H^{4}k\pi^{2}\phi^{4}\omega\coth\big[\frac{2kn\pi^{2}\phi^{2}}{H^{2}\omega}\big]\sinh\big[\frac{2k\pi^{2}\phi^{2}}{H^{2}\omega}\big]\bigg)\bigg\}.
\end{split}
\end{equation}
By combining two equations ($\ref{49}$) and ($\ref{51}$), we can obtain solutions for H like the previous section, and we select the positive answer . Thus, a reconstructed relationship for potential leads to creating new relations that can be used to continue the calculations. In the continuation of the computational process in this section, we want to benefit the $\omega-\omega'$, which is a powerful tool to determine different dark energy models.
 Therefore, we can analyze $\omega-\omega'$ according to all the above  mentioned points in this paper and also using equations $\frac{d\Omega_{D}}{d\ln a}$, ($\ref{41}$), ($\ref{42}$) and ($\ref{43}$).
Therefore, due to the high computational volume, we mention the result according to the free parameters mentioned in the text by plotting a figure.
 We will analyze the results and compare them with the previous model.  Furthermore, we will limit our calculations by considering the complex part of the scalar field and taking the real part of the quintessence field to be a "slow-rolling" field. So, according to the equations ($\ref{10}$), ($\ref{11}$) and ($\ref{43}$), we get the following equation
\begin{equation}\label{52}
\begin{split}
&\frac{\omega^{2}}{a^{6}+\phi^{2}}=\frac{1}{2}\big(\dot{\phi}^{2}+\frac{\omega^{2}}{a^{6}\phi^{2}}\big)=\frac{3c^{2}\phi^{2}H^{4}}{4\omega k}\sinh\big(\frac{2k\pi^{2}\phi^{2}}{\omega H^{2}}\big)\\
&\times\bigg[1+\bigg(-1-\frac{2n}{3}-\frac{4k\pi^{2}\phi^{2}}{3\omega H^{2}}\coth\big[\frac{2kn\pi^{2}\phi^{2}}{\omega H^{2}}\big]-\frac{4}{3}\big(1-\frac{4k\pi^{2}\phi^{2}}{3\omega H^{2}}\coth\big[\frac{2k\pi^{2}\phi^{2}}{\omega H^{2}}\big]\big)\frac{\dot{H}}{H^{2}}\bigg)\bigg].
\end{split}
\end{equation}

Different solutions are obtained for the scalar field $\phi$ that we just  keep only positive solutions.
Given the relation $\dot{\phi}=H\frac{d\phi}{d\ln a}$, the parameter $H$ can not be ignored. So we set the $\frac{d\phi}{d\ln a}$ to zero, i.e., $\frac{d\phi}{d\ln a}=0$.
This relation is formed precisely according to the same process we did in the previous section.
Therefore, the final expression for this case is expressed in the following form.
As it is clear from the following relation, a link is established between the free parameters of the model and $\Omega_{D}$. So,  by setting these free parameters, we can obtain the $\Omega_{D}$, which it  can be compared by the latest observable data,
\begin{equation}\label{53}
\begin{split}
&+\frac{k^{2}\omega\Omega_{D}\big[54(\Omega_{0}^{m}H_{0}^{2}\varphi_{0}^{2}+\omega)-\big\{8n(-3+2n(-3+n\omega))+9(6-5n)\Omega_{D}\big\}\big]}{\sqrt{1+\frac{k^{2}}{c^{4}}}}\\
&-9c^{2}(6-5n)\omega\sqrt{1+\frac{k^{2}}{c^{4}}}=0.
\end{split}
\end{equation}

........As we explained in the previous section, there is a relationship between the free parameters for KHDE in BD cosmology and $ \Omega_{D} $ ($\ref{53}$).
 Concerning mentioned free parameters in the text of this paper, two different values for parameter $\Omega_ {D}$ are always calculated, one of which is within the allowable range so that it can describe the real universe.
 Answers that are outside of the  allowable range, ie $\Omega_{D}>1$, as the previous model, can not fully describe the real universe in non-interacting case , so it is necessary to examine interacting case.
 Note that by setting $\Omega_{D}=0.73$, we can also calculate the value of each of the required free parameters. Of course, in the interactive part, by adding the coupling parameters of the mentioned model, a similar analysis can be carried out, and new results can be proposed, which we will mention in detail in the next section.
\begin{figure}[h!]
 \begin{center}
 \subfigure[]{
 \includegraphics[height=6cm,width=6cm]{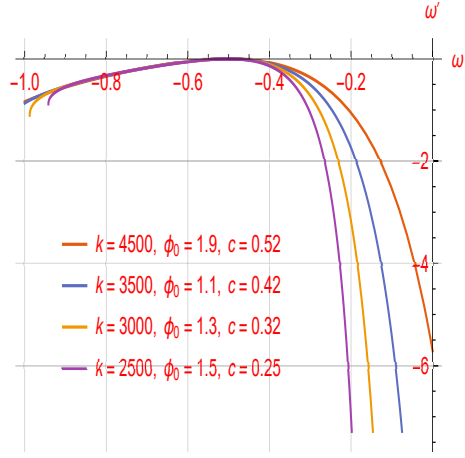}
 \label{4a}}
  \caption{\small{$\omega_{D}-\omega_{D}'$ analyze for the non-interacting case and concerning constant parameters $\omega=10$, $n=0.005$, $k=0$, $\Omega_{m}^{0}=0.3$, $H_{0}=67.9$, and various values of $\phi_{0}, k, c$  }}
 \label{4}
 \end{center}
 \end{figure}

As it is shown in figure ($\ref{4}$), the free parameters $k$, $c$ and $\phi_{0}$  for the non-interacting case are determined by analyzing $\omega_{D}-\omega_{D}'$  for KHDE  in framework of BD.
 For $\omega_{D}<-0.4$, the figures approaching each other show some an adaptation and overlap.
$\omega_{D}'$ also tends to zero for $-0.6<\omega_{D}<-0.4$, and a maximum is observed in this range.
As we see here, the results are very different from the interacting and non-interacting cases of the previous model (Tsallis). Of course, by resetting the free parameters, a similar figure is finally obtained.
 There are some differences and similarities with the interacting case mentioned in the next section,  which we will explain in detail.
It should be noted that the range of free parameters mentioned in this article has been selected according to \cite{71}.
\subsection{Interacting case}

Now as a final part, we are going to investigate KHDE in BD cosmology with interacting case. So, according to equation ($\ref{45}$) and for simplify, we will have

\begin{equation}\label{54}
\begin{split}
\mathcal{N}=-\omega_{D}=-\bigg\{-1+\frac{1}{3}(\alpha-\beta-2n)-\frac{4k\pi^{2}\phi^{2}}{3\omega H^{2}}\coth\big[\frac{2kn\pi^{2}\phi^{2}}{\omega H^{2}}\big]\big(n-\frac{\dot{H}}{H^{2}}\big)+\frac{k\alpha(2\omega n^{2}-2n-1)}{9c^{2}H^{2}\sinh\big(\frac{2kn\pi^{2}\phi^{2}}{\omega H^{2}}\big)}\bigg\}.
\end{split}
\end{equation}

According to the mentioned explanations as well as the relation $\omega_{D}=\omega_{\phi}$, one can obtain,

\begin{equation}\label{55}
\begin{split}
\frac{\mathcal{M}-V(\phi)}{\mathcal{M}+V(\phi)}=-\mathcal{N}
\end{split}
\end{equation}

and

\begin{equation}\label{56}
\begin{split}
V(\phi)=-\frac{\mathcal{N}+1}{\mathcal{N}-1}\mathcal{M}
\end{split}
\end{equation}

Combining two equations ($\ref{49}$) and ($\ref{56}$), yields the following equation

\begin{equation}\label{57}
\begin{split}
-\frac{\mathcal{N}+1}{\mathcal{N}-1}\mathcal{M}=\frac{3c^{2}\phi^{2}H^{4}}{4\omega k}\sinh\big(\frac{2k\pi^{2}\phi^{2}}{\omega H^{2}}\big)-\mathcal{M}.
\end{split}
\end{equation}

By solving the above equation, we can calculate different solutions for the parameter $H$, which can be used for the computational process.
However, the important point here is the condition of self-consistency, for which the parameter $\mathcal{N}$ must be less than 1. In this way, we can obtain a constraint on the potential concerning equations ($\ref{7}$) and ($\ref{49}$).
We also proceed to the analysis $\omega-\omega'$ as in the previous sections. According to the relation $\dot{\Omega}_{D}=H\Omega'_{D}$ and the equations ($\ref{45}$) and ($\ref{46}$), the structure $\omega-\omega'$ can be formed.
As in the previous subsection, we will analyze the result by plotting some figures according to the free parameters of the mentioned model, and we will describe the results in detail.
In addition, for the final task, i.e., considering of the complex part of the scalar field and taking the real part of the quintessence field to be a "slow-rolling" field, according to equations ($\ref{10}$), ($\ref{11}$), ($\ref{49}$) and ($\ref{48}$) we will have.

\begin{equation}\label{58}
\begin{split}
&\frac{\omega^{2}}{\phi^{2}+a^{6}}=\frac{3c^{2}\phi^{2}H^{4}}{4\omega k}\sinh\big(\frac{2k\pi^{2}\phi^{2}}{\omega H^{2}}\big)\\
&\times\bigg[1+\bigg(-1+\frac{1}{3}(\alpha-\beta-2n)-\frac{4k\pi^{2}\phi^{2}}{3\omega H^{2}}\coth\big[\frac{2kn\pi^{2}\phi^{2}}{\omega H^{2}}\big]\big(n-\frac{\dot{H}}{H^{2}}\big)+\frac{k\alpha(2\omega n^{2}-2n-1)}{9c^{2}H^{2}\sinh\big(\frac{2kn\pi^{2}\phi^{2}}{\omega H^{2}}\big)}\bigg)\bigg].
\end{split}
\end{equation}

Solving the above equation, several solutions are obtained for the scalar field which we consider only the positive answers.
Also, concerning relation $\dot{\phi}=H\frac{d\phi}{d\ln a}$, and taking into account that one cannot ignore $H$, leads to $\frac{d\phi}{d\ln a}=0$.
Hence, after calculations, simplification and adjustment of free parameters, the final result for the interacting case of KHDE  in the BD framework is expressed in the following form
\begin{equation}\label{59}
\begin{split}
&4\sqrt{1+\frac{k^{2}}{c^{4}}}(1+n)\pi^{2}\Omega_{m}^{0^{2}}H_{0}^{2}\Omega_{D}+2\omega^{2}(2n(-3+n\omega)-3\Omega_{m}^{0^{2}}+6\Omega_{D})\\
&+\sqrt{1+\frac{k^{2}}{c^{4}}}\omega^{2}(-3+2n^{2}\omega-6n\Omega_{m}^{0^{2}}+16\Omega_{D})\times\big[\alpha-2n(-1+n\omega)\alpha\\
&-3\alpha\Omega_{D}+3(-6+2n+\beta)\Omega_{D}\big]=0.
\end{split}
\end{equation}

As it is clear from the above relation, there is a deep relationship between the free parameters of the mentioned model and $\Omega_{D}$ which can be obtained by setting each of these parameters to acceptable values for $\Omega_{D}$.
As we know, $\Omega_{D}$ can not be greater than 1.
Thus, a close relationship is created between these free parameters and $\Omega_{D}$, which can lead to exciting results.
Therefore, the values of these parameters are significant. We will calculate $\Omega_{D}$ according to these parameters and compare the results for both models with the latest observable data.
As we explained in the previous section, there is a relationship between the free parameters for KHDE in BD cosmology and $\Omega_{D}$, as is evident in the equation ($ \ref{59} $).
 According to the free parameters mentioned in the text of this paper, two different values are obtained for the $ \Omega_{D} $. Of course, we can add that by setting $ \Omega_ {D} = 0.73 $; we can also calculate the value of each of the required free parameters as $\alpha$ and $\beta$.
On the other hand, in this section, by setting the free parameters, we will examine a different structure and calculate different values for the $\Omega_ {D}$.
Therefore, we will have;\\\\
$\bullet\Omega_{m}^{0}=0.3, H_{0}=69.7, n=0.06, \phi_{0}=0.4, c=0.9, \omega=10, k=0.8, \alpha=0.15, \beta=0.25 \longmapsto \Omega_ {D}=0.0054,  \Omega_ {D}=0.8454$\\\\
$\bullet\Omega_{m}^{0}=0.3, H_{0}=69.7, n=0.06, \phi_{0}=0.4, c=1.1, \omega=10, k=0.8, \alpha=-0.25, \beta=-0.2 \longmapsto \Omega_ {D}=0.00066, \Omega_ {D}=0.7750$ \\\\
$\bullet\Omega_{m}^{0}=0.3, H_{0}=69.7, n=0.06, \phi_{0}=0.4, c=1.2, \omega=10, k=0.8, \alpha=-0.2, \beta=0.15 \longmapsto  \Omega_ {D}=0.0015,\Omega_ {D}=0.8186$\\\\
$\bullet\Omega_{m}^{0}=0.3, H_{0}=69.7, n=0.06, \phi_{0}=0.4, c=1.3, \omega=10, k=0.8, \alpha=0.15, \beta=0.25 \longmapsto \Omega_ {D}=0.0063,\Omega_ {D}=0.8896$.\\

As we have mentioned, we can determine each of these parameters in different ways to be within its acceptable range. In this section, we have carried out a further analysis that can be repeated for previous examples. A noteworthy point in the whole of this paper, which is of great importance, is that a dynamic model has been studied for the expansion of the universe in a dynamical framework, i.e., BD, which can have better results and be closer to the observable data.

\begin{figure}[h!]
 \begin{center}
 \subfigure[]{
 \includegraphics[height=6cm,width=6cm]{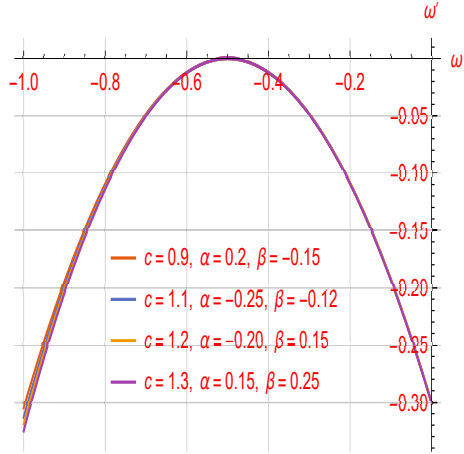}
 \label{5a}}
  \caption{\small{$\omega_{D}-\omega_{D}'$ analyze for the interacting case and concerning constant parameters $\omega=10$, $n=0.06$, $k=0$, $\Omega_{m}^{0}=0.3$, $H_{0}=67.9$, $k=0.8$, $\phi_{0}=0.4$, and various values of $\beta, \alpha, c$  }}
 \label{5}
 \end{center}
 \end{figure}

For the interacting case of the KHDE in BD cosmology, as shown, we plot  figure ($ \ref{5} $) to obtain the  $ \omega_{D}-\omega_{D}'$ concerning the free parameters mentioned in the text.
Figures have negative values and for different free parameters, especially $c$ and coupling parameters ($\alpha, \beta$) in the interacting sample, the figures have very close values, which seem to show a complete overlap.
 There is a maximum in $\omega_{D}=-0.5$, and $\omega_{D}'$ tends to be zero at this point.
Of course, as the parameters ($\omega_{D}$) and ($\omega_{D}'$) get smaller, the figures get further apart, and their overlap and compliance decrease.
As mentioned before, the coupling parameters play a vital role in this model.
Definitely, the results of this case are somewhat different from the non-interacting case, especially in the narrowing of the figures and the overlap.
\section{Summary and  concluding remarks}
The primary purpose of this paper was to study two different structures of holographic dark energy, namely Tsallis and Kaniadakis in the context of Brans-Dicke cosmology, considering the complex form of the quintessence model. Hence, we used the non-interacting and interacting cases of the above two models to calculate some cosmological parameters such as the equation of state ($\omega$).
In this regard, we performed the $ \omega- \omega '$ analysis for the models mentioned in both the non-interacting and interacting cases. We examined the complex quintessence cosmology by modifying the potential and studying the scalar field dynamics.
In addition, considering the two parts of the quintessence field effects, i.e., real and complex, and considering the fractional energy density $ \Omega_{D} $, we examined whether it can describe a real universe or not.
We also specified that the fractional energy density could not be arbitrary between 0 and 1. In other words, it depends on the free parameters defined in both holographic dark energy models, namely Tsallis, Kaniadakis, and Brans-Dicke cosmology.
We obtained a relationship between the fractional energy density and other parameters introduced in the text, such as $\delta$, $b^{2}$, $\alpha$, and $\beta$, for each model separately. Finally, we compared the results of models with respect to each other and the latest observable data.

 All the above explanations can raise some fundamental questions here.\\
It is possible to study the complex form of other scalar fields such as phantom field, correspond with different dark energy models, and compare them with the results obtained from this paper.\\
Also can be examined for holographic dark energy, and the results can be compared with this article.
Different structures of dark energy can be studied in different contexts and systems as fractal universe and higher dimensional and challenged by corresponding to the complex part of the scalar field.\\
Other models of holographic dark energy arising from super statistics entropy such as Renyi and Sharma-Mittal holographic dark energy can be challenged according to the structure of this paper and compared with the latest observable data.

\end{document}